\documentclass[preprint,12pt]{elsarticle}

\usepackage{graphicx}
\usepackage{bm}
\usepackage{dcolumn}
\usepackage{amsmath}
\usepackage{amssymb}

\begin{document}

\begin{frontmatter}

\title{Magnetotransport properties of FeSe in fields up to 50~T. }
\author[l1]{Y.A.~Ovchenkov\corref{cor1}}
\ead{ovtchenkov@mig.phys.msu.ru}

\author[l2,l3,l4]{D.A.~Chareev}
\author[l1]{V.A.~Kulbachinskii}
\author[l1]{V.G.~Kytin}
\author[l1,l5]{D.E.~Presnov}
\author[l6]{Y.~Skourski}
\author[l1,l3,l7]{O.S.~Volkova}
\author[l1,l7,l8]{A.N.~Vasiliev}

\address[l1]{Faculty of Physics, M.V. Lomonosov Moscow State University, Moscow 119991, Russia}
\address[l2]{Institute of Experimental Mineralogy RAS, Chernogolovka, Moscow District, 142432, Russia}
\address[l3]{Ural Federal University, 620002 Ekaterinburg, Russia}
\address[l4]{Kazan Federal University, 18 Kremlyovskaya Str., Kazan, 420008, Russia }
\address[l5]{Skobeltsyn Institute of Nuclear Physics, Moscow 119991, Russia}
\address[l6]{Dresden High Magnetic Field Laboratory (HLD-EMFL), HZDR,  Dresden, Germany}
\address[l7]{National University of Science and Technology “MISiS,” 119049 Moscow, Russia}
\address[l8]{National Research South Ural State University, 454080 Chelyabinsk, Russia}

 \cortext[cor1]{Corresponding author}

\begin{abstract}
Magnetotransport properties of the high-quality FeSe crystal, measured in a wide temperature range and in magnetic fields up to 50~T, show the symmetry of the main holelike and electronlike bands in this compound.     
In addition to the main two bands, there is also a tiny, highly mobile, electronlike band which is responsible for the non-linear behavior of $\rho_{xy}(B)$ at low temperatures and some other peculiarities of FeSe. We observe the inversion of the $\rho_{xx}$ temperature coefficient at a magnetic field higher than about 20~T which is an implicit conformation of the electron-hole symmetry in the main bands.

\end{abstract}


\end{frontmatter}

\section{Introduction}
FeSe is a very important and interesting superconducting material with complicated electronic and transport properties \cite{2017_Coldea}. It is a nearly compensated semimetal with low carrier concentration. For the physics of superconductivity, it is a new type of superconducting materials and it is a new playground to test out the existing theories of superconductivity. In particular, the low carrier concentration should have allowed a significant variation of a superconducting transition temperature ($T_{c}$) under variation of a carrier concentration. Indeed, it is demonstrated that the transition temperature can be substantially varied using a gate electrode \cite{PhysRevLett.116.077002}. However the  pairing mechanism in FeSe and other iron-based superconductors is still being debated, and the reasons, causing a $T_{c}$ increase under pressure \cite{Medvedev2009},  and for a mono-layer FeSe film on an epitaxial substrate \cite{CPL-29-3-037402}, are unclear.

The properties of FeSe, as well as many other iron-based superconductors, can not be described by a simple two-band semimetal model. The first studies of the iron-based superconductors revealed multiband effects and electron-hole asymmetry in Ba(FeCo)${}_{2}$As${}_{2}$  \cite{PhysRevB.80.140508}. Later, an analysis of the magnetic field dependence of $\rho_{xy}$ and $\rho_{xx}$ suggested the presence of the highly mobile electronlike band in BaFe${}_{2}$As${}_{2}$ \cite{PhysRevB.84.184514}. The similar highly mobile band exists in many other iron-based superconductors including FeSe family \cite{SUST-30-3-035017} and, apparently, originates from a small local region of the Fermi surface. Since the mobilities of the two main bands are several times lower than for the highly mobile band, their properties can be studied separately in a high magnetic field where the conductivity of the highly mobile band is suppressed significantly.    
   
Here we report the magnetotransport properties of the high-quality FeSe crystal measured in a wide temperature range and magnetic fields up to 50~T. The obtained data prove a good symmetry of the main electronlike and holelike bands. A remarkable phenomenon is observed at temperatures below 100~K.  All $\rho_{xx}(B)$ curves, corresponding to different temperatures, cross each other in the region 15-20~T and 0.1-0.15 m$\Omega$cm. Therefore, a crossover from a metallic-type $\rho_{xx}(T)$ to a semiconductor-type  dependence occurs at a magnetic field higher than 20~T. Such behavior has a simple description within the two-band model which gives another way to extract the parameters of the main bands.

%
\section{Experiment}
The FeSe crystals were grown using the KCl/AlCl${}_{3}$ flux technique \cite{CrystEngComm12.1989}. 
The chemical composition of the crystals was studied with the energy dispersive micro analysis system. The composition measurements were done at three points for four average size crystals.

Electrical measurements were done on a cleaved rectangular sample with lengths of 1.2~mm, widths 0.6~mm and thicknesses about 0.05~mm. Contacts were made by sputtering of Au/Ti layers through a precisely machined mechanical mask. The current electrodes were 0.1~mm wide lines along small sides of the bar. Potential and Hall 0.1$\times$0.1 mm$^{2}$ electrodes were connected to a sample holder with a 0.025~mm gold wire using H20E silver epoxy.  

DC magnetoresistance and Hall effect measurements were done using EDC options of Quantum Design MPMS 7T with Keithley 2400 and Keithley 2182A. Measurements of resistance in pulsed magnetic fields up to 50~T were done in HLD at HZDR, Germany.

\section{Results and discussion}
\begin{figure}[ht]
\includegraphics[scale=0.5,angle=0]{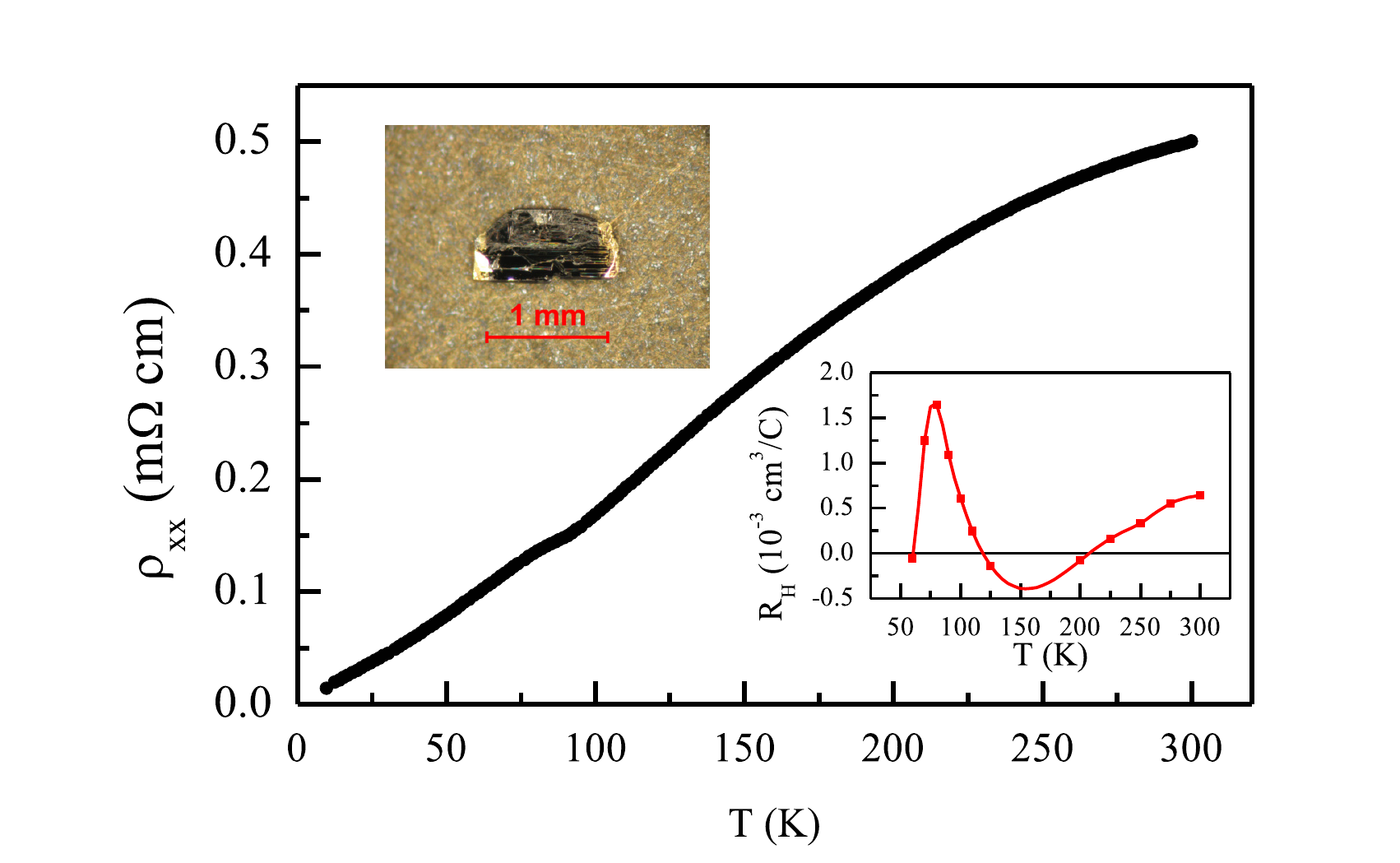}
\caption{\label{fig1} The temperature dependence of the resistivity  $\rho_{xx}$. Top inset: the picture of the investigated crystal. Bottom inset: Temperature dependence of the Hall coefficient $R_{H}$.  }
\end{figure}
The temperature dependence of the sample resistance is shown in Fig.1. The anomaly at $\rho_{xx}(T)$ around 90~K corresponds to the structural phase transition. The resistivity at 15~K is 22 times lower than at 300~K which indicates a high quality of used crystal. An optical image of the studied crystal is shown in the top inset of Fig.1.The bottom inset of Fig.1 shows the temperature dependence of the Hall coefficient $R_{H}$. The Hall coefficient has sign reversal points in presented temperature range which is one of the consequences of a carrier compensation. The low-temperature behavior of $R_{H}$ is not plotted because of non-linearity of $\rho_{xy}(B)$ at low temperatures.
\begin{figure}[ht]
\includegraphics[scale=0.5,angle=0]{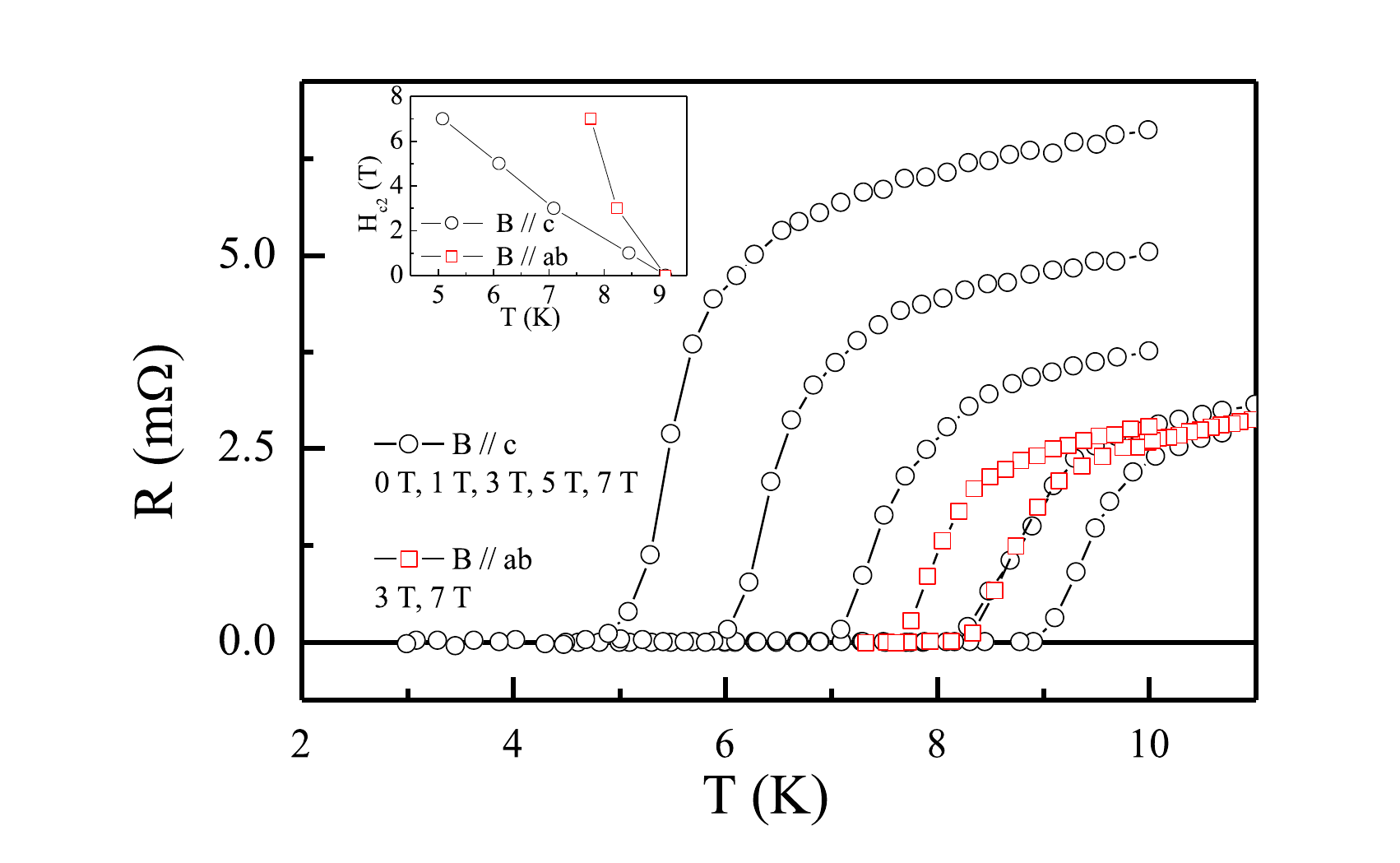}
\caption{\label{fig2} The temperature dependence of the resistance near a superconducting transition at the various magnetic fields. Inset: Temperature dependence of H${}_{C2}$ critical field for the two field orientations.  }
\end{figure}
The quality of the sample is also confirmed by $R(T,B)$ measurements near the superconducting transition temperature. Figure 2 shows $R(T)$, measured around the transition, in magnetic fields parallel and perpendicular to the crystal plane. The inset shows the temperature dependencies of the critical fields for these two orientations determined at zero-resistivity points. The ratio of slopes for these dependencies is 3.1-3.4. It is the ratio of coherence lengths for the $ab$ plane and for the $c$ axis direction. This value of anisotropy is close to the highest reported for FeSe single crystals which confirms a perfect layered structure of the studied crystal.

The field dependence of the resistivity tensor components within a quasiclassical relaxation-time approximation can be expressed as a sum of $l$ band terms: 
\begin{eqnarray}
\sigma_{xx}=\sum_{i=1}^{l}\frac{\sigma_{i}}{(1+\mu_{i}^{2}B^2{})}\\
\sigma_{xy}=\sum_{i=1}^{l}\frac{s_{i} \sigma_{i}\mu_{i}B}{(1+\mu_{i}^{2}B^2{})}\\
\sigma_{i}=en_{i}\mu_{i}
\end{eqnarray}
where $\sigma_{xx}$ and $\sigma_{xy}$  are conductivity tensor components, $i$ is a band index,  $\sigma_{i}$,  $\mu_{i}$, and $n_{i}$ are absolute values of a band conductivity, a carrier mobility and a concentration correspondingly; $s_{i}$ is "-1" for a hole and "+1" for an electron bands.
Resistivity tensor components $\rho_{xx}$ and $\rho_{xy}$  in a tetragonal crystal are related as follows: 

\begin{eqnarray}
\rho_{xx}=\rho_{yy}=\frac{\sigma_{xx}}{(\sigma_{xx}^{2}+\sigma_{xy}^{2})} \\
-\rho_{xy}=\rho_{yx}=\frac{\sigma_{xy}}{(\sigma_{xx}^{2}+\sigma_{xy}^{2})}
\end{eqnarray}

\begin{figure}[ht]
\includegraphics[scale=0.5,angle=0]{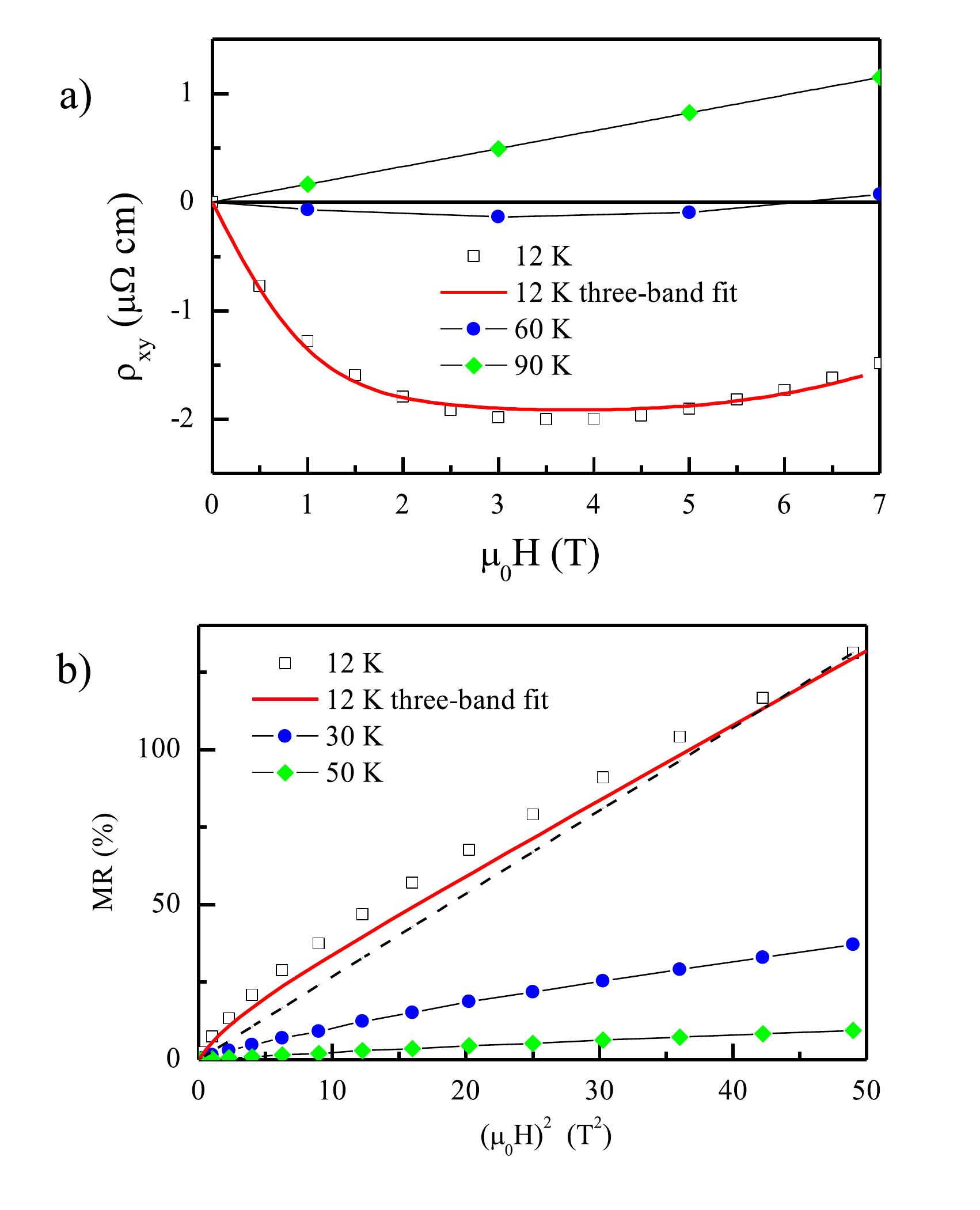}
\caption{\label{fig3} (a) The field dependence of $\rho_{xy}$. (b) Magnetoresistance  $MR=({\rho_{xy}(B)-\rho_{xy}(0)})/{\rho_{xy}(0)}$ versus $B^{2}$. The straight dash line in (b) highlights the curvature of the 12~K curve. The experimental data at 12~K are fitted with the three-band model.  }
\end{figure}

For the two-band material, this model gives a linear in $B$ law for $\rho_{xy}$ and $B^{2}$ for $\rho_{xx}$ in the limit $\mu_{i}B \ll 1$ for both mobilities. The measured dependencies $\rho_{xy}(B)$  and magnetoresistance $MR(B)=(\rho_{xx}(B)-\rho_{xx}(0))/\rho_{xx}(0)$ in dependence on $B^{2}$ are shown in Fig. 3 a) and b) respectively. It is clearly seen that the curves corresponding 12~K deviate substantially from a linear dependence. A similar behavior was reported for many iron-based superconductors and can be described within the three-band model by adding to a couple of main electron and hole bands, which have almost the same concentrations and mobilities, the tiny band with appreciably higher mobility \cite{SUST-30-3-035017}. The Table 1 lists the results of the simultaneous fit of $\sigma_{xy}(B)$ and $\sigma_{xx}(B)$ data, obtained at 12~K and 30~K, with three-band equations (1) - (3). The ratios  $n_{e}/n_{h}$ and $\mu_{e}/\mu_{h}$ at 12~K are 0.84 and 1.12 respectively. Therefore, in a quasiclassical relaxation-time approximation FeSe can be described as having the two main electron and hole bands, with approximately equal values of a concentration and a mobility, and a tiny mobile band with a 3-5 \% of the total carrier concentration which provide about 10-15 \% of the total conductivity in zero fields. The relative contribution to the total conductivity of this highly mobile band rapidly decreases with increasing magnetic field. For example, accordingly the data listed in Table 1 the band $e_{2}$ provides 13\% of the total conductivity at 12~K in zero field and only about 2\% in 5~T.

\begin{table*}[h]
\caption{\label{Table1}  The results of the simultaneous fitting of $\sigma_{xy}(B)$ and $\sigma_{xx}(B)$ in field range up to 7 T using the three-band model. }
\begin{tabular}{ccccc}
 \hline
 & \multicolumn{2}{c}{12 K}&\multicolumn{2}{c}{30 K}\\
$band$   & $n$ & $\mu$ & $n$ & $\mu$\\
 & (10$^{19}$ cm$^{-3}$) & (cm$^{2}$/Vs) & (10$^{19}$ cm${}^{-3}$) & (cm$^{2}$/Vs)\\
 \hline
$e_{1}$&8.98&1537&8.4&670 \\
$h_{1}$&10.67&1365&10.5&615 \\
$e_{2}$&0.66&6481&0.6&2500\\
 \hline
\end{tabular}
\end{table*}

Consequentially, in high magnetic fields a two-band semimetal is a good model to describe the FeSe transport properties. It allows to give a simple explanation for a crossover to a negative temperature coefficient of $\rho_{xx}$ in high magnetic fields which is demonstrated in Fig. 4. This figure shows $\rho_{xx}(B)$ corresponding to temperatures in the range 30-80~K. All curves cross each other in the field range 15-25~T. Therefore, the temperature coefficient of resistivity $\rho_{xx}$ changes it's sign from a positive in low magnetic fields to a negative in high magnetic fields. The inset of Fig.4 shows $\rho_{xx}(T)$ at zero and 45~T magnetic fields which confirm this inversion.
\begin{figure}[ht]
\includegraphics[scale=0.5,angle=0]{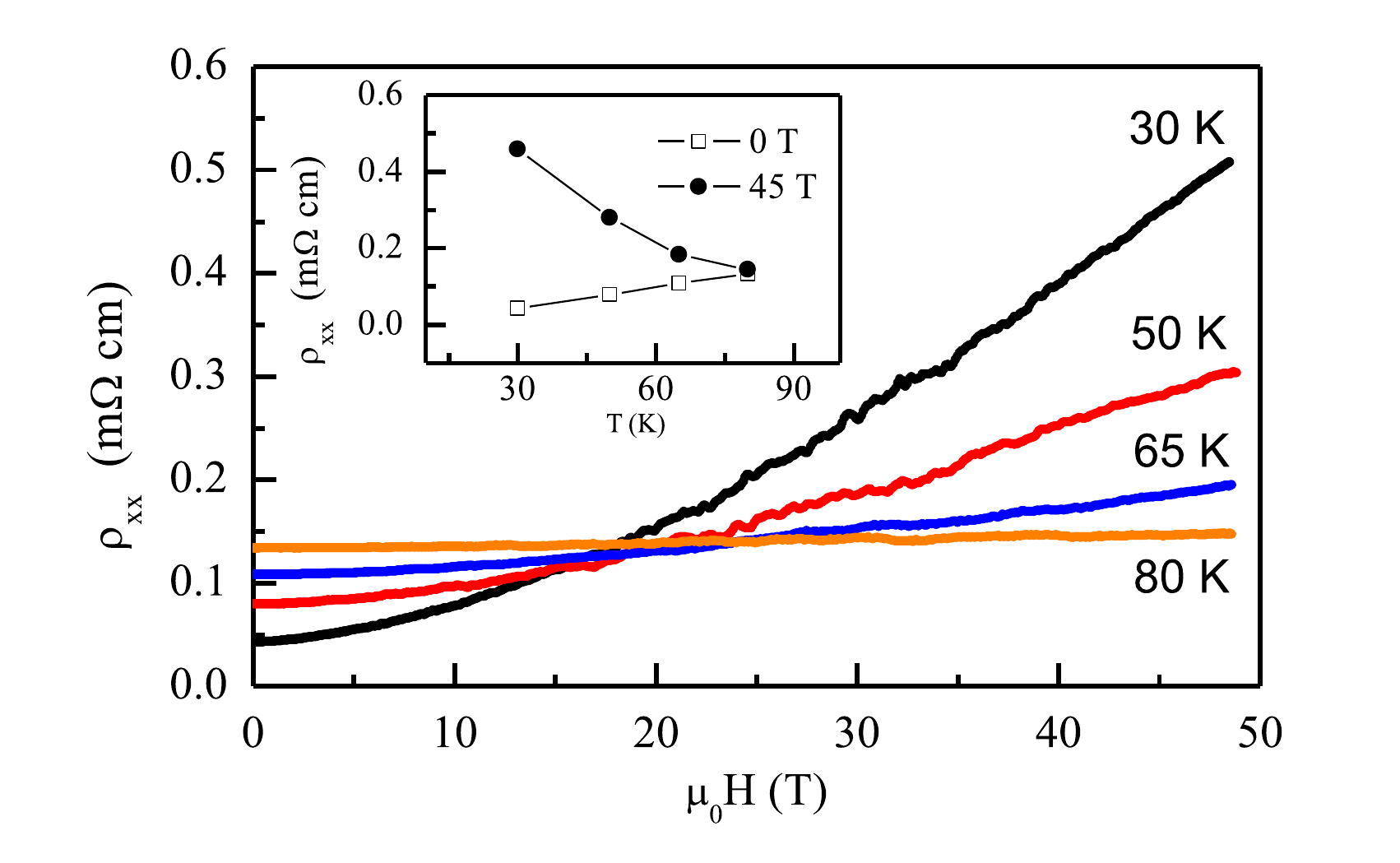}
\caption{\label{fig4} Field dependence of the resistivity  $\rho_{xx}$ at temperatures from 30 to 80 K. Inset : The temperature dependence of $\rho_{xx}$ at 0 and 45~T. }
\end{figure}

To get a clear and comprehensive expression, we consider the simplest possible two-band model with the equal values of the electron and hole concentrations and mobilities: $n_{e}=n_{h}=n_{0}/2$, $\mu_{e}=\mu_{h}=\mu_{0}$, and consequently  $\sigma_{e}=\sigma_{h}=\sigma_{0}/2$.
Substitution in (1) and (2) gives:
\begin{eqnarray}
\sigma_{xx}=\sum_{i=e,h}\frac{\sigma_{i}}{(1+\mu_{i}^{2}B^2{})}
=\frac{\sigma_{0}}{(1+\mu_{0}^{2}B^2{})}\\
\sigma_{xy}=\sum_{i=e,h}\frac{s_{i}\sigma_{i}\mu_{i}B}{(1+\mu_{i}^{2}B^2{})}=0
\end{eqnarray}
Eq. (4) reduces to: 
\begin{eqnarray}  
\rho_{xx}=1/\sigma_{xx}=\frac{1}{\sigma_{0}}(1+\mu_{0}^{2}B^2{}) ;
\end{eqnarray}
therefore MR has simple expression:
\begin{eqnarray}
MR=\mu_{0}^{2}B^2{}
\end{eqnarray}

Eq. (8) can be written as follows:
\begin{eqnarray}
\rho_{xx}=\frac{1}{en_{0}\mu_{0}}+ \frac{\mu_{0}}{en_{0}}B^2{} 
\end{eqnarray}
and we can introduce:
\begin{eqnarray}
\rho_{0}=\frac{1}{en_{0}\mu_{0}}=\frac{1}{en_{0}}\mu_{0}^{-1}\\
\rho_{B}= \frac{\mu_{0}}{en_{0}}B^2{}=\frac{B^2{}}{en_{0}}\mu_{0} 
\end{eqnarray}
where $\rho_{0}$ and $\rho_{B}$ are independent and dependent on magnetic field terms respectively. It is clearly seen that temperature dependencies of these terms, which are due to temperature dependence of $\mu_{0}$, are reciprocal. Therefore, if in a zero magnetic field  $\rho_{xx}(T)= f(T)$, where $f(x)$ - any arbitrary function, then in a strong enough magnetic field $B$:
\begin{eqnarray}
 \rho_{xx}(T)\approx (\frac{B}{en_{0}})^{2}f^{-1}(T); 
\end{eqnarray}

This simple and interesting phenomenon is exactly what observed for FeSe. It allows to find the carrier concentration from the relation:
\begin{eqnarray}
(\rho_{xx}(B)-\rho_{xx}(0))\rho_{xx}(0)=(\frac{B}{en_{0}})^{2}
\end{eqnarray}
  
If $\mu_{1}$ and $\mu_{2}$ are the mobility values at temperatures $T_{1}$ and $T_{2}$, then intersection at $B_{x}$ gives the next equation:
\begin{eqnarray}
\frac{1}{en_{0}\mu_{1}}+ \frac{\mu_{1}}{en_{0}}B_{x}^{2} =\frac{1}{en_{0}\mu_{2}}+ \frac{\mu_{2}}{en_{0}}B_{x}^{2}
\end{eqnarray}
The solution of Eq. (15) is:
\begin{eqnarray}
B_{x}^2{}=\frac{1}{\mu_{1}\mu_{2}}
\end{eqnarray}

Using obtained relations we can determine "the experimental" values of the band's parameters. At 30~K $\rho_{xx}(0)$=4.28$\times$10$^{-5}$ $\Omega$cm and $\rho_{xx}(45 T)$=4.59$\times$10$^{-4}$ $\Omega$cm. For these values the equation (14) gives 2.16$\times$10$^{20}$ cm$^{-3}$ for carrier concentration and (9) gives 600 (cm$^{2}$/Vs) for carrier mobility. These values are in a good agreement the values listed in Table 1. From equation (16) it follows that  $\rho_{xx}(B)$ curves for the temperature range corresponding the mobility range [$\mu_{1}, \mu_{2}$] will cross each other in the field range [$1/\mu_{1}, 1/\mu_{2}$]. It gives again about 600-700 (cm$^{2}$/Vs) for the mobility value at 30~K.   In general, it shows that the magnetotransport properties of FeSe can be satisfactory quantitatively described within the quasiclassical approximation. Reported violation of the Kohler's rule for these compounds \cite{PhysRevB.93.180503} probably related to the variation of the highly mobile band parameters.

The data in Table 1 show a good symmetry of the main electron and hole band parameters. These data where obtained from the transport measurements up to fields satisfying relation $\mu B \approx 1$ for the main bands. To improve the accuracy of comparison and to probe the carriers properties in higher fields we fitted the $\rho_{xx}(B)$ dependence measured at 1.5~K (see Fig. 5) with explicit two band expression for MR \cite{Kim-jap84}:
\begin{eqnarray}
MR=\frac{\alpha B^{2}}{(1+\beta B^{2})}\\
\alpha=\sigma_{e}\sigma_{h} (\mu_{e} +\mu_{h} )^2 /(\sigma_{e} +\sigma_{h} )^{2}\\
\beta = ({\mu_{e}\mu_{n} ( n_{e} - n_{h})})^{2}/( \sigma_{e}+\sigma_{h})^{2}
\end{eqnarray}
where $\alpha$ is a slope of  $\rho_{xx}(B^{2})$ in low fields and $\beta $ describes a saturation of $\rho_{xx}(B^{2})$. The ratio $\beta / \alpha$ is:
\begin{eqnarray}
\frac{\beta}{\alpha}=\frac{\mu_{e}\mu_{h}}{(\mu_{e}+\mu_{h})^2}\frac{( n_{e} - n_{h})^{2}}{n_{e}n_{h}}
\end{eqnarray}
  The coefficient $\beta$ allows to determine an exact value of the non-compensation $ \lvert (n_{e} -n_{h})\rvert$.

\begin{figure}[ht]
\includegraphics[scale=0.5,angle=0]{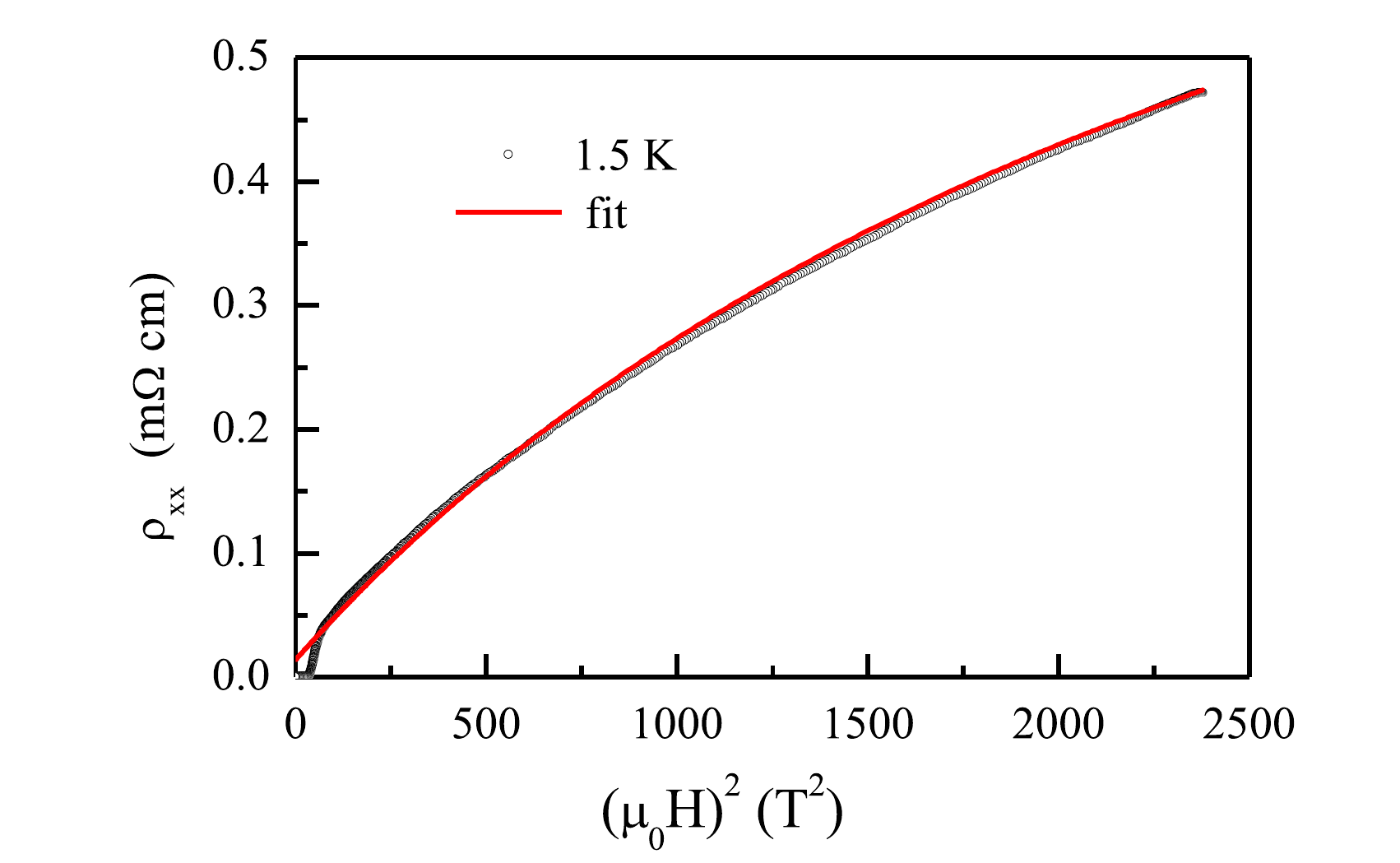}
\caption{\label{fig4} Experimental dependence of $\rho_{xx}$ at 1.5~K and fitting function $\rho_{1.5} \times (1+MR)$ for $\rho_{1.5}$=1.42$\times$10$^{-5}$ $\Omega$cm and MR expressed in Eq. (17) with $\alpha$=2.43$\times$10$^{-2}$ (T$^{-2}$), and $\beta$ =3.3$\times$10$^{-4}$ (T$^{-2}$). }
\end{figure}

For the almost symmetrical $n_{e} \approx n_{h}=n_{0}/2$, $\mu_{e} \approx \mu_{h}=\mu_{0}$, and $\sigma_{e} \approx \sigma_{h}=\sigma_{0}/2$. Then $\alpha \approx \mu_{e}\mu_{n} \approx \mu_{0}^{2}$ and $\beta / \alpha \approx (( n_{e} - n_{h})/ n_{0})^{2}$ . Then the slope of  $\rho_{xx}(B^{2})$, which is equal to 2.43$\times$10$^{-2}$ (T$^{-2}$), gives for mobility values, if, suppose, they are equal, about 1560 (cm$^{2}$/Vs). It is the same value as it were determined for the main bands at 12~K (see Table 1) which can be explained by a mobility saturation at low temperatures. The ratio $\beta / \alpha$ is 1.36$\times$10$^{-2}$ then $\lvert (n_{e} -  n_{h})\rvert)/ n_{0}$ is about 0.11 which gives $(n_{e} / n_{h}$ = 1.22 or 0.82. The last value is in a good agreement with 0.84 which was previously obtained for data in Table 1.

\section{Conclusion}
The experimental data on the transport properties of the high-quality FeSe crystal in high magnetic fields show a good symmetry of the main electron and hole bands. In particular, this symmetry allows observing the inversion of the temperature coefficient of resistivity. The properties of the tiny highly mobile band deserve a further investigation because of it can be responsible for the measurable interaction of ultrasound with electrons observed in FeSe \cite{epl_0295-5075-101-5-56005} and, therefore, can play a special role in an appearance of superconductivity in this compound.

\section{Acknowledgments}
This work was supported by the Ministry of Education and Science of the Russian Federation in the framework of Increase Competitiveness Program of NUST “MISiS” project K2-2016-066, by the Russian Government Program of Competitive Growth of Kazan Federal University and by Act 211 of  Government of the Russian Federation, contracts 02.A03.21.0004, 02.A03.21.0006 and 02.A03.21.0011. We acknowledge support of Russian Foundation for Basic Research through Grants ofi-m-16-29-03266, 15-03-99628. We acknowledge the support of HLD at HZDR, a member of the European Magnetic Field Laboratory (EMFL).

 \bibliographystyle{elsarticle-num} 
 \bibliography{FeSe-50T.bib}

\listoffigures

\end{document}